\documentclass[conference]{IEEEtran}
\IEEEoverridecommandlockouts
\usepackage{newtxmath}

\usepackage[dvipsnames]{xcolor}
\usepackage{tabularx}
\usepackage{multirow}
\usepackage{float}
\usepackage{amsmath}
\usepackage{url}
\usepackage{xcolor,colortbl}
\usepackage[shortlabels]{enumitem}
\usepackage{mathtools}
\usepackage{multirow}
\usepackage{bm}
\usepackage{tikz-cd}
\usepackage{float}
\usepackage{caption}
\usepackage[labelfont=bf]{caption}
\usepackage{subcaption}
\usepackage{graphicx}
\usepackage{makecell}
\definecolor{salmon}{rgb}{0.945,0.784,0.631}
\definecolor{flarered}{rgb}{0.945,0.251,0.275}
\definecolor{grannysmith}{rgb}{0.659,0.894,0.627}

\newcommand\bvec{\mathbf}

\usepackage{xcolor,pifont}
\newcommand*\colourcheck[1]{%
  \expandafter\newcommand\csname #1check\endcsname{\textcolor{#1}{\ding{52}}}%
}
\newcommand*\colourcross[1]{%
  \expandafter\newcommand\csname #1cross\endcsname{\textcolor{#1}{\ding{55}}}%
}
\colourcheck{green}
\colourcross{red}

\usetikzlibrary{matrix,arrows}
\usepackage[version=4]{mhchem}
\usepackage{upgreek}
\usepackage{verbatim}
\usepackage{physics}
\usepackage{dcolumn}
\usepackage{titlesec}

\usepackage{pgfplots,pgfplotstable}
\usepgfplotslibrary{groupplots}
\usepgfplotslibrary{colorbrewer}
\pgfplotsset{cycle list/Set1}
\usepgfplotslibrary{fillbetween}
\pgfplotsset{compat=1.8}
\pgfplotsset{
    tick align=outside,
    tick pos=left,
    xmajorgrids,
    x grid style={white},
    ymajorgrids,
    y grid style={white},
    axis line style={white},
    axis background/.style={fill=white!89.803921568627459!black},
    legend style={draw=none, fill=none},
    legend cell align=left,
}
\pgfkeys{/pgf/number format/.cd, 1000 sep={\,}}

\pgfplotsset{
    log x ticks with fixed point/.style={
        xticklabel={
            \pgfkeys{/pgf/fpu=true}
            \pgfmathparse{2^\tick}%
            \pgfmathprintnumber[fixed relative, precision=4]{\pgfmathresult}
            \pgfkeys{/pgf/fpu=false}
        }
    },
    log10 x ticks with fixed point/.style={
        xticklabel={
            \pgfkeys{/pgf/fpu=true}
            \pgfmathparse{10^\tick}%
            \pgfmathprintnumber[fixed relative, precision=3]{\pgfmathresult}
            \pgfkeys{/pgf/fpu=false}
        }
    },
    log y ticks with fixed point/.style={
        yticklabel={
            \pgfkeys{/pgf/fpu=true}
            \pgfmathparse{2^\tick}%
            \pgfmathprintnumber[fixed relative, precision=4]{\pgfmathresult}
            \pgfkeys{/pgf/fpu=false}
        }
    }
}
\tikzset{font=\small}

\titlespacing{\section}{0pt}{\parskip}{-\parskip}
\titlespacing{\subsection}{0pt}{\parskip}{-\parskip}
\titlespacing{\subsubsection}{0pt}{\parskip}{-\parskip}

\graphicspath{ {./fig/} }

\begin{document}

\title{Scaling the leading accuracy of deep equivariant models to  biomolecular simulations of realistic size}

\makeatletter
\newcommand{\linebreakand}{%
  \end{@IEEEauthorhalign}
  \hfill\mbox{}\par
  \mbox{}\hfill\begin{@IEEEauthorhalign}
}
\makeatother

\author{
\linebreakand
\IEEEauthorblockN{Albert Musaelian$^*$, Anders Johansson$^*$, Simon Batzner$^*$}
\IEEEauthorblockA{
\{albym,andersjohansson,batzner\}@g.harvard.edu}
\textit{Harvard John A. Paulson School of} \\
\textit{Engineering and Applied Sciences,} \\
Cambridge, MA USA\\
\and
\IEEEauthorblockN{Boris Kozinsky$\dagger$}
\IEEEauthorblockA{
bkoz@seas.harvard.edu\\
\textit{Harvard John A. Paulson School of} \\
\textit{Engineering and Applied Sciences,} \\}
\IEEEauthorblockA{
\textit{Robert Bosch Research and Technology Center}\\
Cambridge, MA USA\\
}
}
\maketitle
\thispagestyle{plain}
\pagestyle{plain}

\def\thefootnote{*}\footnotetext{Equal contribution. Order is random.}\def\thefootnote{\arabic{footnote}}

\def\thefootnote{$\dagger$}\footnotetext{Corresponding author: \url{bkoz@seas.harvard.edu}\\ }\def\thefootnote{\arabic{footnote}}

\begin{abstract}
This work brings the leading accuracy, sample efficiency, and robustness of deep equivariant neural networks to the extreme computational scale. This is achieved through a combination of innovative model architecture, massive parallelization, and models and implementations optimized for efficient GPU utilization. The resulting Allegro architecture bridges the accuracy-speed tradeoff of atomistic simulations and enables description of dynamics in structures of unprecedented complexity at quantum fidelity. To illustrate the scalability of Allegro, we perform nanoseconds-long stable simulations of protein dynamics and scale up to a 44-million atom structure of a complete, all-atom, explicitly solvated HIV capsid on the Perlmutter supercomputer. We demonstrate excellent strong scaling up to 100 million atoms and 70\% weak scaling to 5120 A100 GPUs.
\end{abstract}

\section{Justification}
First scalable, transferable machine-learning potential with state-of-the-art equivariant deep-learning accuracy. Performance of 100 timesteps/s for range of biomolecular systems. 70\% weak scaling to 1280 nodes and 5120 A100 GPUs, excellent strong scaling up to 100 million atoms. First application of state-of-the-art machine learning interatomic potentials to large-scale biomolecular simulations.
\section{Performance Attributes}
\begin{table}[H]
\begin{center}
\begin{tabularx}{\columnwidth}{l|l}
\hline
Performance Attribute & Our Submission \\
\hline
\hline
Category of achievement & Scalability, time-to-solution\\
\hline
Performance & 100 timesteps/s \\
\hline
Maximum problem size & 126.4 million atoms\\
\hline
\multirow{2}{*}{Type of method used} & Explicit (molecular dynamics, Allegro \\ &  equivariant deep learning potentials)  \\
\hline
Results reported on basis of & Whole application including I/O \\
\hline
Precision reported & Mixed precision (with GPU tensor cores)\\
\hline
\multirow{2}{*}{System scale} & Full-scale system \\
& 1280 nodes (5120 GPUs)\\
\hline
Measurement mechanism & Wall time, timesteps/s \\
\hline
\end{tabularx}
\end{center}
\end{table}

\section{Problem Overview: First-Principles Dynamics of Matter \label{sec:problem}}

The ability to predict the time evolution of matter on the atomic scale is the foundation of modern computational biology, chemistry, and materials engineering. Even as quantum mechanics governs the microscopic atom-electron interactions in vibrations, migration and bond dissociation, phenomena governing observable physical and chemical processes often occur at much larger length- and longer time-scales than those of atomic motion. Bridging these scales requires both innovation in fast and highly accurate computational approaches capturing the quantum interactions and in extremely parallelizable architectures able to access exascale computers. 
Presently, realistic physical and chemical systems are far more structurally complex than what computational methods are capable of investigating, and their observable evolution is beyond the timescales of atomistic simulations. This gap between key fundamental questions and phenomena that can be effectively modeled has persisted for decades. From one side of the gap, models of small size, representing ostensibly important parts of the systems, can be constructed and investigated with high-fidelity computationally expensive models, such as electronic structure methods of density functional theory (DFT) and wave-function quantum chemistry. In the domain of materials science, these models can capture individual interfaces in metallic composites, defects in semiconductors, and flat surfaces of catalysts. However, evolution of such structures over relevant time scales is  out of reach with electronic structure methods. Importantly, such reduction of complexity is not possible in the domain of biological sciences, where entire structures of viruses consist of millions of atoms, in addition to similarly large number of explicit water molecules needed to capture the physiological environment. From the other side of the gap, uncontrolled approximations have to be made to reach large sizes and sufficient computational speeds. These approximations have relied on very simple analytical models for interatomic interactions and have many documented failures of describing dynamics of both complex inorganic and biological materials \cite{robustelli2018developing}.

Molecular dynamics (MD) simulations are a pillar of computational science, enabling insights into the dynamics of molecules and materials at the atomic scale. MD provides a level of resolution, understanding, and control that experiments often cannot provide, thereby serving as an extremely powerful tool to advance our unstanding and design of novel molecules and materials. Molecular dynamics simulates the time evolution of atoms according to Newton's equations of motion. By integrating the forces at each time step, a sequence of many-atom configurations is obtained, from which physical observables can then be obtained. The bottleneck of MD is the short integration time step required, which usually lies on the order of femtoseconds. Since many chemical and biological processes occur on the timescale of microseconds or even milliseconds, billions to trillions of integration steps are needed. This highlights the requirement common to all MD simulation for access to atomic forces in a way that is simultaneously both a) accurate and b) computationally efficient. For decades, two different avenues have been pursued: on one hand, classical force-fields (FFs) are able to simulate large-scale systems with high computational efficiency, enabling simulations of billions of atoms and reaching even microsecond simulation time-scales on special-purpose hardware \cite{Anton2}. The simple functional form of classical FFs, however, greatly limits their predictive power and results in them not being able to capture quantum-mechanical effects or complex chemical transformations. On the other hand, quantum-mechanical simulations provide a highly accurate description of the electronic structure of molecules and materials. Interatomic forces computed with these first-principles methods can then be used to integrate the atoms in molecular dynamics, called \textit{ab-initio} molecular dynamics (AIMD), with DFT being the most common method of choice. The cubic scaling of plane-wave DFT with the number of electrons, however, is the key bottleneck limiting AIMD simulations as it severly limits both length- and time-scales of AIMD simulations, only allowing thousands of atoms and hundreds of picoseconds in routine simulations.

\section{Current State of the Art\label{sec:sota}}
\subsection{Accuracy}
Over the past two decades, machine learning interatomic potentials (MLIPs) have been pursued with immense interest as a third approach, promising to bypass this long-standing dilemma \cite{behler2007generalized, bartok2010gaussian, zhang2018deep, thompson2015spectral, schutt2018schnet, batzner20223, musaelian2023learning}. The aim of MLIPs is to learn energies and forces from high-accuracy reference data while scaling \emph{linearly} with the number of atoms. Initial efforts combined hand-crafted descriptors \cite{behler2007generalized, bartok2010gaussian} with a Gaussian Process or a shallow neural network. These first MLIPs were quickly further improved \cite{zhang2018deep, drautz2019atomic, thompson2015spectral} and have been adapted to biomolecular simulations, trained on large data sets to provide general-purpose potentials \cite{ani-1, ani-2}. Despite this initial progress, however, this first generation of MLIPs has been severely limited in predictive accuracy, often unable to generalize to structures different from those in the training data, resulting in simulations that are not transferable and often not robust \cite{batzner20223, kovacs2021linear, musaelian2023learning, fu2022forces}. In an effort to overcome these limitations in accuracy, more recently, deep learning interatomic potentials based on the message passing neural network (MPNN) paradigm have been proposed and shown to be a more accurate alternative to the first generation of MLIPs, however at a significant computational overhead \cite{schutt2018schnet}.

Common to all interatomic potentials is a focus on symmetry: in particular, the energy must obey invariance with respect to translations, rotations, and reflections, which together comprise E(3), the Euclidean group in 3D. Both classical force-fields and modern MLIPs achieve this by only ever acting on geometric invariants of the underlying structures. More recently, in an attempt to better represent the symmetry of the data, \emph{invariant} interatomic potentials have been generalized to \emph{equivariant} ones \cite{batzner20223, schutt2021equivariant, batatia2022design, batatia2022mace, musaelian2023learning}. In equivariant MLIPs, the hidden network features consist of not only scalar features, but also vectors and higher-order tensors, resulting in a more faithful representation of the atomistic geometry. The initial NequIP work on equivariant interatomic potentials demonstrated not only a step change in state-of-the-art accuracy, but also the ability to outperform the invariant DeepMD accuracy while training on a thousand times smaller data set \cite{batzner20223}. Following these efforts, equivariant MLIPs have been shown to greatly improve stability of simulations \cite{fu2022forces} and display much better extrapolative power than existing approaches \cite{batzner20223, batatia2022design, batatia2022mace, musaelian2023learning}. 
\begin{figure*}
    \centering
    \includegraphics[width=0.8\textwidth]{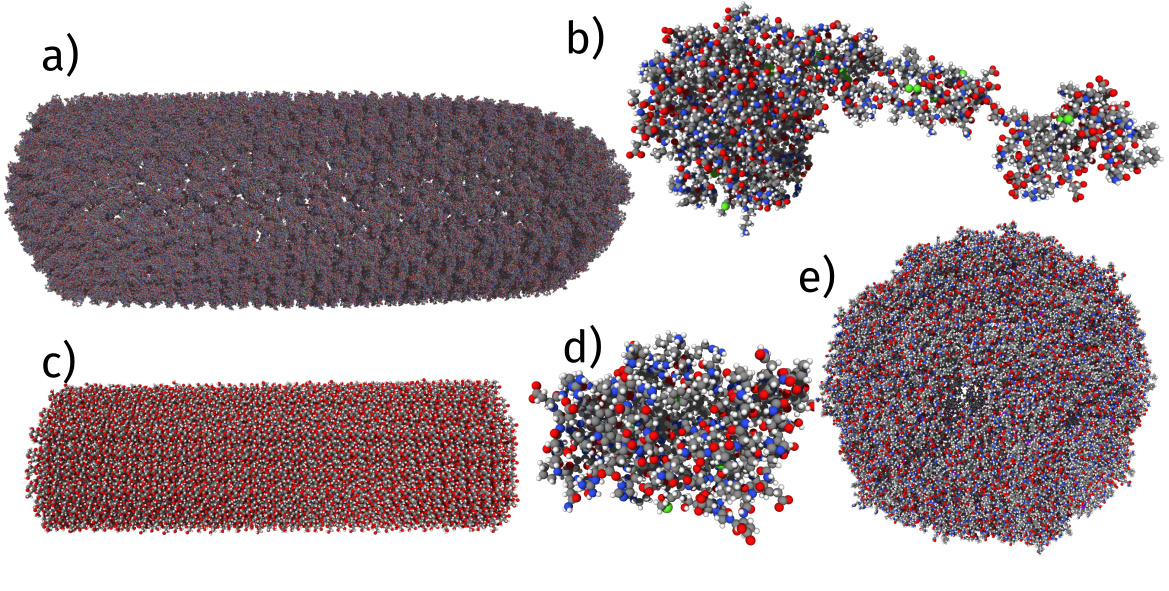}
    \caption{Biomolecular systems used for benchmarking: a) HIV capsid, 44M atoms b) factor IX, 91k atoms c) cellulose, 409k atoms d) DHFR, 23k atoms e) STMV, 1M atoms. The solvating water has been omitted for visualization.}
    \label{fig:biosystems}
\end{figure*}
Equivariant potentials however, struggle with scale and speed since all existing methods combine the equivariance with the MPNN architecture of previous generation methods. This is due to the graph propagation mechanism in MPNNs: at each layer of message passing, information is passed from a central node to its neighbors. While interactions between a central atom $i$ and its neighbors $j$ are \emph{local} to a finite-range interaction sphere (typically between 4-6~\AA), each iteration of the MPNN increases this receptive field. As an illustrative example, \cite{musaelian2023learning} chose a system of bulk water: while for a local cutoff of 6~\AA, each atom has on average 96 neighbors, for a typical six-layer MPNN, this receptive field grows to 36~\AA\ and now includes 20,834 atoms. This propagation mechanism makes message-passing potentials near-impossible to scale across multiple devices and multiple nodes. Thus the total number of interacting neighbors for an atom grows \emph{cubically} with the number of layers in the neural network. In order to scale a message passing network, one would have to either maintain enormous neighbor lists, or transfer messages and gradients between devices at each layer. Both of these approaches are destined to be slow, with the latter also requiring extensive software development to allow the neural network's message passing to work together with the MD software's spatial decomposition message passing. Thus the exceptional accuracy of equivariant methods remains inaccessible for applications that require large length scales and long timescale simulations, with biological systems being a major example that is virtually entirely excluded from these state-of-the-art approaches.

\begin{table*}[!htb]
\caption{Comparison of different interatomic potentials, best method in bold, second best underlined. \textbf{Left:} results on the internal energy $U_0$ in the QM9 benchmark, measured by the mean absolute error (MAE) in [meV]. Allegro outperforms all existing message-passing and transformer-based architectures while being the \emph{only method} able to scale. \textbf{Right:} MAE in forces on rMD17, units of [meV/\AA], averaged over all molecules with models trained on a per-molecule basis. Existing local methods perform significantly worse than state-of-the-art equivariant potentials. All existing equivariant potentials however do not scale, with the exception of Allegro. For details on the methods, see \cite{musaelian2023learning, kovacs2021linear}.}
    \begin{minipage}{.4\linewidth}
      \centering
        \begin{tabular}{lcc}
\hline \hline
Model & $U_0$ & Strictly Local \\
\hline
Cormorant \cite{anderson2019cormorant}& 22 & \redcross \\
SchNet \cite{schutt2018schnet}     & 14 & \redcross \\ 
EGNN \cite{satorras2021n} & 11 & \redcross \\
DimeNet++ \cite{gasteiger2020fast} & 6.3 & \redcross \\
SphereNet \cite{liu2021spherical}& 6.3 & \redcross \\
ET \cite{tholke2022torchmd} & 6.2 & \redcross \\
PaiNN \cite{schutt2021equivariant} & 5.9 & \redcross \\ \hline \hline
Allegro, 1 layer & \underline{5.7 (0.3)} & \greencheck \\
Allegro, 3 layers & \textbf{4.7} (0.2) & \greencheck  \\
\hline
    \end{tabular}
    \end{minipage}%
    \begin{minipage}{.6\linewidth}
      \centering
        \begin{tabular}{lccc}
\hline \hline
Model & MAE, F & Equivariant & Strictly Local\\
\hline
Classical force-field \cite{wang2004development} & 227.2 &  No & \greencheck + Coulomb \\ \hline
ANI-random \cite{ani-1, ani-2} & 50.71 & No & \greencheck \\ 
ANI-pretrained \cite{ani-1, ani-2}   & 25.89 & No & \greencheck \\ 
GAP \cite{bartok2010gaussian} & 22.54  & No & \greencheck \\
ACE \cite{drautz2019atomic} & 8.79 & No & \greencheck \\  \hline
sGDML \cite{sgdml} & 14.48 & No & \redcross \\ \hline
OrbNet-Equi \cite{qiao2021unite} & 4.31 & Yes & \redcross \\
NequIP \cite{batzner20223}  & 3.52 & Yes & \redcross \\ 
MACE \cite{batatia2022mace}  & \underline{2.92} & Yes  & \redcross \\ \hline  \hline
Allegro  & \textbf{2.81 }& Yes & \greencheck \\ 
\hline
    \end{tabular}
    \end{minipage} 
    \label{tab:mol-perf}
\end{table*}

\subsection{Scalability and speed}
In parallel to the improvements in accuracy, great strides have been achieved in terms of the computational cost of MLIPs, with several methods sacrificing some accuracy in order to be able to perform extreme-scale simulations. Most notably, the DeePMD method won the 2020 Gordon Bell supercomputing award for their 100 million-atom simulations, using the entire Summit machine \cite{zhang2018deep, jia2020pushing}. The following year, SNAP extended the scale to 20 billion atoms, accompanied by a 20-fold increase in speed \cite{nguyen2021billion}. Finally, FLARE set the current record for GPU-accelerated MLIP uncertainty-aware reactive MD benchmarks with 0.5 trillion atoms on Summit and a 70\% speed increase over SNAP \cite{johansson2022micron}. 

\subsection{Quantum-accurate biomolecular simulations}

\begin{table}[htbp]
\centering
    \caption{Sample efficiency of the equivariant Allegro potential: RMSE of forces on liquid water and three ices in units of [meV/\AA].}
    \label{tab:deepmd-water}
\begin{tabular}{lccc}
\hline \hline
 &   Allegro  & DeepMD  \\
\hline 
$N_\text{train}$ &  133 & 133,500\\ 
\hline 
Liquid Water  & \textbf{29.1}  & 40.4 \\          
Ice Ih (b)    & \textbf{30.7} & 43.3 \\                           
Ice Ih (c)    & \textbf{21.0} & 26.8 \\           
Ice Ih (d)    &  \textbf{18.0} & 25.4\\  \hline
Strictly Local                 & \greencheck & \greencheck \\

  \hline \hline
    \end{tabular}
\end{table}
Owing to the large length- and time-scales required for MD simulations in many biological applications, empirical potentials such as the AMBER force-field \cite{wang2004development} remain dominant and widely used. Due to the limitations of classical FFs, however, there is strong interest in increasing the accuracy of biomolecular simulations. Deep learning interatomic potentials have been applied to biomolecular systems in hopes of achieving higher accuracy \cite{unke2022accurate}, but only 25k atom scale has been reached due to the lack of scalability of MPNNs \cite{SpookyNet}. In parallel, hybrid approaches have been explored that only treat the solute-solute interactions with the MLIP, while solvent-solvent and solvent-solute interactions were modelled with a classical polarizable force-field \cite{jaffrelot2023scalable}. Hence, these approaches lose the high accuracy advantages available in princieple with the advanced MLIPs. Efforts to scale to larger biomolecular systems at high accuracy have been led so far not by machine learning but instead by quantum approaches. In particular, extreme-scale linear-scaling ab-initio DFT quantum calculations have been conducted on millions of atoms (summarized in \cite{schade2022towards}), and semi-empirical tight-binding calculations have been scaled up to tens of millions of atoms (see table \ref{tab:water-strong-comparison}).
We note that, although difficult to compare directly, given the combination of DFT reference data using a high-performance hybrid functional and the high accuracy of the Allegro model, it is likely that the MLIP is already more accurate than the existing large-scale semi-empirical quantum calculations. Further, the accuracy of the MLIP can in principle be improved systematically with higher-quality reference calculations while maintaining its speed and scaling.
\begin{table}[htbp]
    \centering
    \caption{Comparison with previous effort towards large-scale quantum accuracy effort, demonstrating \({>}1000\times\) improvement in time-to-solution. Speed from strong scaling of approximately 1M atom water simulations, see figure~\ref{fig:strongscaling} for this work and figure 10 of \cite{schade2022towards}.}
    \label{tab:water-strong-comparison}
    \begin{tabular}{r|c|cccc}
         & \multirow{2}{*}{\# atoms} & \multicolumn{4}{c}{Timesteps/s on \# nodes} \\
        & &  16 & 32 & 64 & 1024\\
        \hline
        Tight binding \cite{schade2022towards} & 1,022,208 & 0.010 & 0.012 & 0.020 & - \\
        This work & 1,119,744 & 6.28 & 11.9 & 20.3 & 104.2 
    \end{tabular}
\end{table}

\section{Innovations Realized \label{sec:innovation}}
\subsection{Allegro: Scalable Equivariant Deep Learning\label{subsec:allegro}}

Allegro \cite{musaelian2023learning} is the first scalable equivariant MLIP to overcome the previous dilemma of choosing between highly accurate, robust, transferable \cite{fu2022forces} equivariant models and scalable, but less accurate, first-generation models.
Allegro attains this by first decomposing the total predicted energy of the system into atomic energies $E_{\text{system}} = \sum_{i}^{N}{\sigma_{Z_i} E_i + \mu_{Z_i}}$  where $\sigma_{Z_i}$ and $\mu_{Z_i}$ denote a per-species scale and shift parameter, respectively. This atomic energy is then further decomposed into a sum of per-ordered-pair energies $E_{i} = \sum_{j \in \mathcal{N}(i)}{ E_{ij}}$. Allegro's equivariant features $\bvec{V}^{ij, L}_{n,\ell,p}$ are thus also indexed by an ordered pair of neighboring atoms $(i, j)$ at each layer $L$. These features formally inhabit a direct sum of irreducible representations (``irreps'') of the $O(3)$ rotation and mirror symmetry group, which are indexed by a rotation order $\ell=0,1,2...$ and parity $p=\pm 1$. Intuitively, they are comprised of scalars ($\ell=0$), vectors ($\ell=1$), and higher-order geometric tensors ($\ell \geq 2$). The $n$ index is an additional feature channel index.

The first key innovation of Allegro is its tensor product layer, which updates the features with information about neighboring atoms' geometry using the ``tensor product of representations,'' a fundamental equivariant operation. 
Specifically, the per-ordered-pair tensor features $\bvec{V}^{ij, L-1}$ are updated by taking their tensor product with a learned weighted sum over the spherical harmonic embeddings $\vec{Y}^{ik}$ of the positions of the neighbors $k$ of the central atom $i$:
\begin{align}
    \bvec{V}^{ij, L}_{n,(\ell_1,p_1, \ell_2,p_2) \to (\ell_\text{out}, p_\text{out})} &= \sum_{k \in \mathcal{N}(i)}{ w_{n, \ell_2, p_2}^{ik,L}  \left( \bvec{V}^{ij, L-1}_{n,\ell_1,p_1} \otimes \vec{Y}^{ik}_{\ell_2,p_2} \right)
    } \\&= 
    \bvec{V}^{ij, L-1}_{n,\ell_1,p_1} \otimes \left( \sum_{k \in \mathcal{N}(i)}{w_{n, \ell_2, p_2}^{ik,L}  \vec{Y}^{ik}_{\ell_2,p_2}} \right)
    \label{eq:tp}
\end{align}
We exploit here the bilinearity of the expensive tensor product operation to avoid computing it for each neighbor $k$, instead summing first over neighbors $k \in \mathcal{N}(i)$ and performing a single tensor product. Because all atom pair indices share the same central atom $i$, information remains \emph{strictly local} to its neighborhood and the growth of the receptive field is \emph{entirely avoided}. This innovation allows Allegro to learn increasingly complex representations of the atomic structure layer after layer while keeping all interactions strictly local, thus making it massively parallelizable.

The second central innovation of Allegro is motivated by the difference in cost between scalar operations and the much more expensive $O(3)$-equivariant operations that are symmetrically permitted on tensors. 
Allegro is therefore designed to put significant network capacity into the scalar operations, particularly dense neural networks, which are comparatively cheap and highly optimized on modern GPU hardware, while keeping as few expensive equivariant operations as possible (only the tensor product) and limiting the number of tensor feature channels. This is achieved by having separate scalar and tensor tracks throughout the network, which at each layer communicate information which allows the high capacity of the scalar track to ``control'' the equivariant features. Across deep learning, it has repeatedly been observed that larger networks perform better \cite{batatia2022design}. In high-performance applications of deep learning such as interatomic potentials, however, this increase in computation is unsustainable as it would hurt computational efficiency. The ability of Allegro to partially decouple these two effects allows it to greatly increase network capacity while keeping the computational overhead moderate. As shown in table \ref{tab:mol-perf}, Allegro outperforms state-of-the-art potentials on energies and forces of small molecules as measured by the QM9 and revMD17 benchmarks, including state-of-the-art equivariant message-passing-based approaches. Remarkably, Allegro also performs significantly better than the DeepMD potential on water despite being trained on more than 1,000 times fewer reference data (see Table \ref{tab:deepmd-water}). This demonstrates that Allegro is able to retain the remarkable improvements demonstrated by equivariant message-passing potentials, while being massively parallelizable, greatly increasing accessible length-scales as well as time-to-solution.

\begin{figure}
    \centering
    \includegraphics[width=\linewidth]{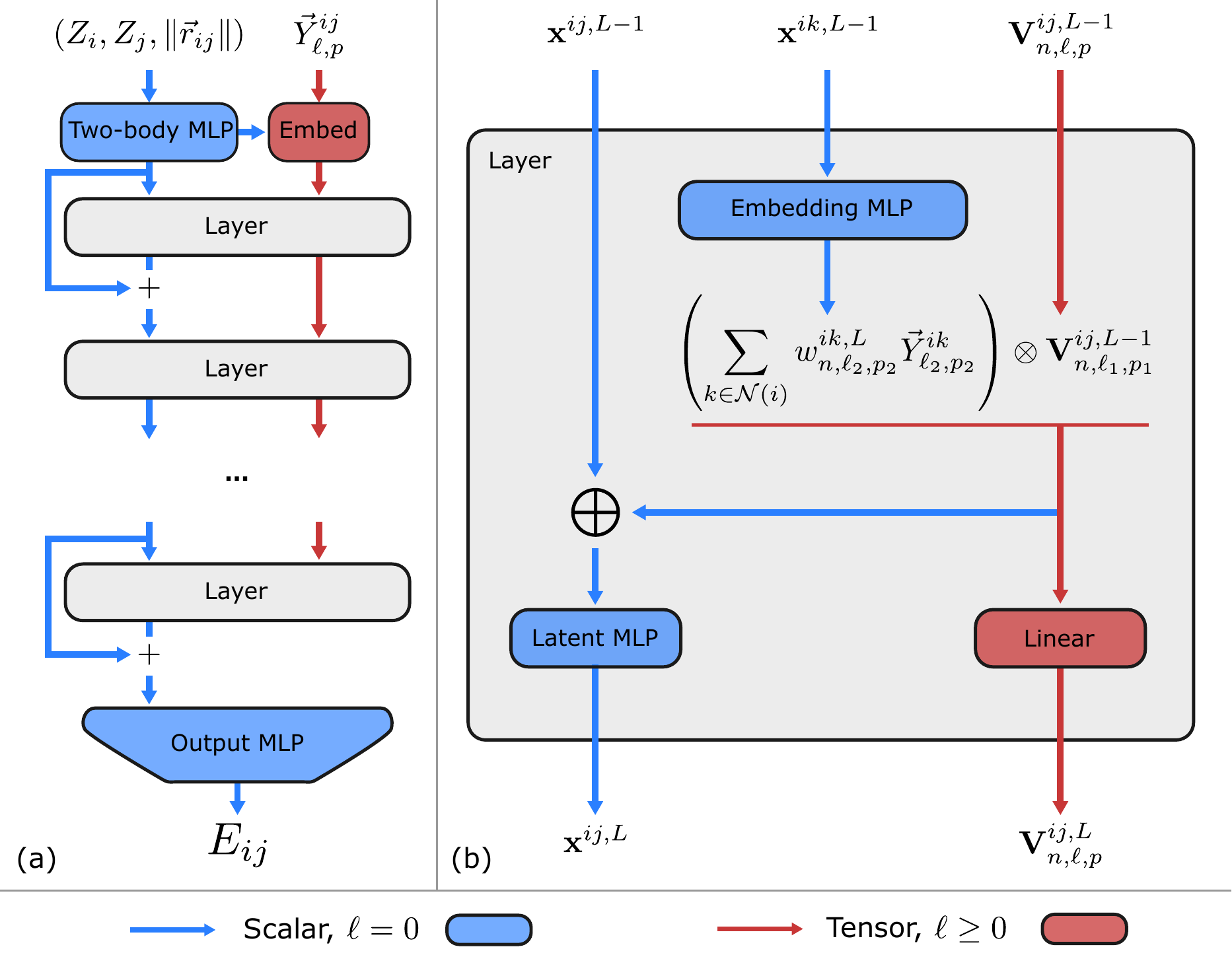}
    \caption{The Allegro network architecture: the network consist of an initial two-body embedding, followed by the core tensor product layer, resulting in the prediction of per-pair energies $E_{ij}$. Scalar (blue) and tensor (red) tracks iteratively share information in an effort to let the high capacity of the computationally cheap scalar track influence the equivariant features.}
    \label{fig:allegro}
\end{figure}

\subsection{Optimization strategies in Allegro\label{subsec:accel}}
\subsubsection{Strided memory layout}
Tensor features of different rotation orders $\ell$ have different dimensions $2\ell + 1$.
A key implementation detail in equivariant neural networks is how best to store groups of such tensors in memory.
Previous equivariant neural networks have either stored tensors of different $(\ell, p)$ in separate arrays, or concatenate the tensors along the $2\ell + 1$ in (arbitrary) order of $\ell$ \cite{geiger2022e3nn}.
For both approaches code size, and thus overhead (particularly on GPUs), scales poorly with $\ell_\text{max}$ as per-$(\ell, p)$ processing/extraction code is necessary.

Our Allegro implementation instead introduces a ``strided layout'' scheme (see \ref{fig:pmodetp}): all tensor features of various $(\ell, p)$ are stored together in a single array whose innermost two dimensions are $[n_\text{tensor}, \sum_{\ell, p}{2\ell + 1} \leq 2 \times (\ell_\text{max} + 1)^2]$. While consuming the same and optimal amount of memory, this layout allows efficient calculations that mix tensors of different $(\ell, p)$ as explained in the next section. Importantly, the use of the strided layout with a homogeneous $n_\text{tensor}$ across irreps is in part made practical by the two-track architecture, which allows the inclusion of arbitrary amounts of scalar capacity in the model without needing to increase $n_\text{tensor}$ for all irreps.

\subsubsection{Optimized tensor product}
\begin{figure*}
    \centering
    \includegraphics[width=\linewidth]{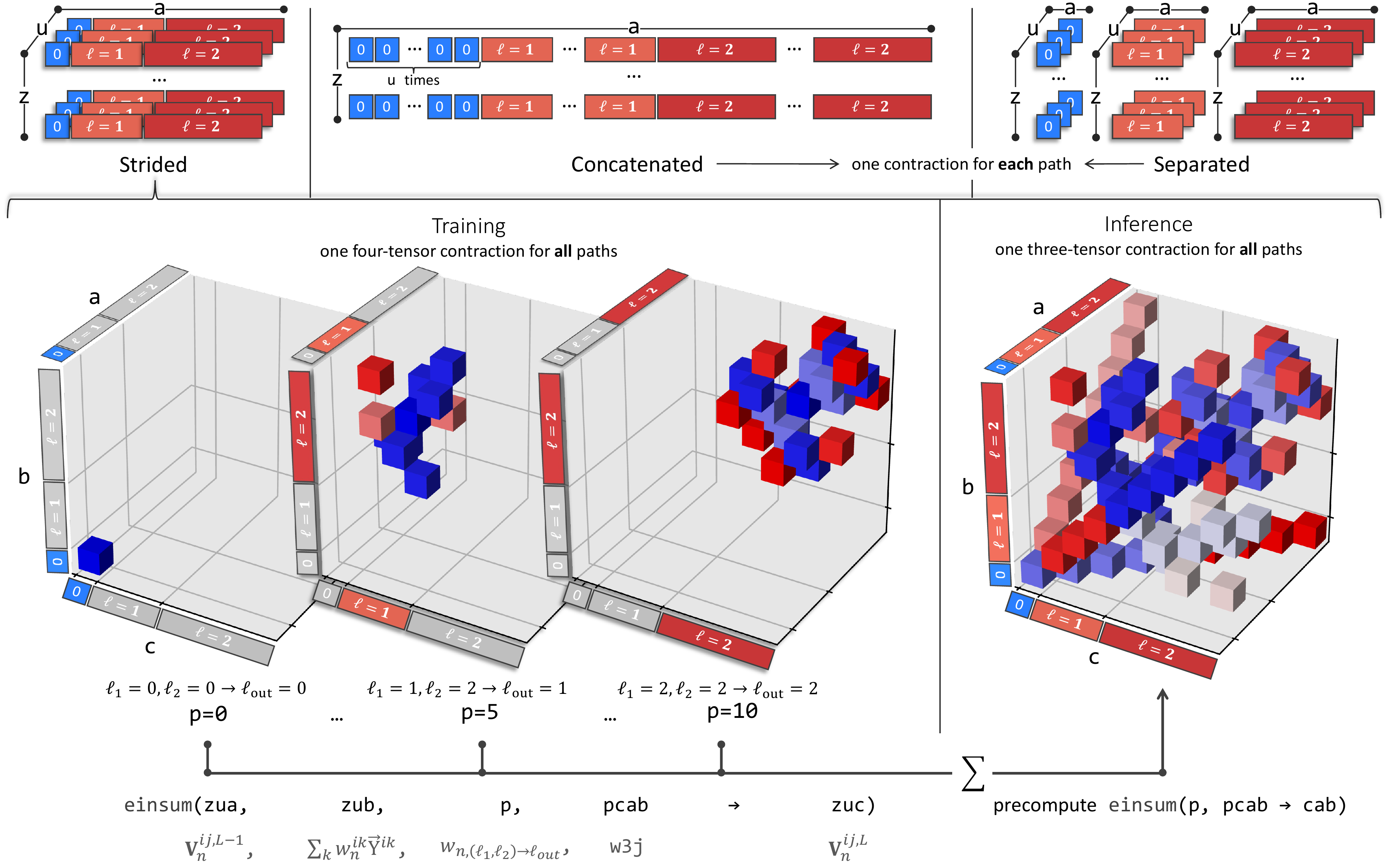}
    \caption{Top: comparison of our strided layout with previous memory layouts for equivariant features. Bottom:  optimized tensor product of strided layouts with precomputed path fusion at inference. Gray cubes show the constant \texttt{w3j} tensor with nonzero Wigner 3j coefficients shown with small blue/red cubes.  The \texttt{a,b,c} indices range over the innermost irrep dimension, and \texttt{p} indexes paths. The \texttt{u} index goes over the $n_\text{tensor}$ tensor feature channels and the \texttt{z} index ranges over neighbor pairs. The operation is also shown in standard Einstein notation with the correspondence to equation \ref{eq:tp} shown.}
    \label{fig:pmodetp}
\end{figure*}

The key equivariant operation in Allegro is the tensor product in equation \ref{eq:tp} which mixes features from different atom pairs \emph{and} tensor features of different irreps $(\ell_1, p_1)$ and $(\ell_2, p_2)$. The tensor product is the most expensive operation in Allegro's tensor track, and the only nonlinearity.

In Allegro we compute all symmetrically valid combinations of input and output irreps, which satisfy $| \ell_1 - \ell_2 | \leq \ell_\text{out} \leq | \ell_1 + \ell_2 |$ and $p_\text{out} = p_1 p_2$. Each such combination is called a ``path'' and can be expressed in Einstein summation notation as the three tensor contraction $\bvec{x}\otimes\bvec{y}_{m_\text{out}} = \text{\texttt{w3j}}_{m_\text{out}, m_1, m_2} \bvec{x}_{m_1} \bvec{y}_{m_2}$ between the input tensors and the constant Wigner 3j symbol denoted \texttt{w3j} \cite{musaelian2023learning}. The number of paths scales unfavorably with $\ell_\text{max}$, which imposes significant overhead and code size on previous efforts that compute them separately. This scaling is improved, but not avoided, by our omitting all tensor product paths that are not symmetrically allowed to eventually contribute to the final scalar outputs.  Due to our strided layout, however, Allegro is able to efficiently pose the entire tensor product as a single tensor contraction whose innermost contraction dimensions are still at largest $2 \times (\ell_\text{max} + 1)^2$, eliminating this per-path overhead. 

In this work we further improve the performance of the tensor product by eliminating the full linear mixture over tensor product paths and feature channels from equation (16) in Allegro \cite{musaelian2023learning}. We replace it with a simpler learned weighted sum over each set of paths sharing the same $(\ell_\text{out}, p_\text{out})$, which we found to have neglible effect on accuracy across a variety of molecular and materials systems. This weighting, unlike the previous full mixture over channels, can be efficiently pre-computed, eliminating the scaling of the tensor product's inference cost with the number of paths (see figure \ref{fig:pmodetp}). The entire tensor product is then a single three tensor contraction that can be decomposed into GPU-efficient dense pairwise (matrix) multiplications. This decomposition is chosen optimally and automatically for any set of hyperparameters using \texttt{opt\_einsum} \cite{daniel2018opt} through the package \texttt{opt\_einsum\_fx} we developed to integrate it with PyTorch's TorchScript compiler.

In the final layer of an Allegro model the only allowable paths are those leading to scalars, for which \texttt{w3j} is only nonzero for $m_1 = m_2$. We further reduce the cost of the tensor product in this case by explicitly removing the redundant dimension from the contraction.

\subsubsection{Mixed-precision calculations}

MLIPs face a difficult trade off with regard to numerical precision: they need to be as fast as possible, which suggests the use of limited or mixed precision, but they are also sensitive regression models for scientific computing where double width (\texttt{float64}) is the norm.
In Allegro, we balance these competing goals with a custom mixed precision setup. For the purposes of training, Allegro models are carefully normalized such that all their weights and internal activation have component-wise magnitudes on the order of 1.0. Consequently, the internal calculations and parameters are well represented using single-precision \texttt{float32} arithmetic. 
For matrix multiplications, which underlie both the latent multi-layer perceptrons (MLPs) and the tensor product, and thus the vast majority of the computational cost, we use NVIDIA's \texttt{TensorFloat32} (TF32), a 19-bit compute format that allows the use of dedicated Tensor Core matrix multiplication processors. 
Finally, we strategically use \texttt{float64} only in the \emph{final} stages of the model to resolve numerical problems resulting from the large magnitudes of typical atomic energies (see, for example, \cite{batatia2022design}) at negligible cost. In particular, we conduct the shifting, scaling, and summation of the atomic energies (see section \ref{subsec:allegro}) in double precision.

\subsubsection{Species-dependent cutoffs}

In chemically diverse systems such as biomolecules, there is significant variation in the distribution of distances between atoms and their neighbors depending on chemical species, and this has been used to reduce the number of neighbor pairs while preserving relevant interactions \cite{johansson2022micron}.

We leverage Allegro's per-ordered-pair architecture to impose such per-species cutoffs at a more granular level:  we use independent cutoffs for each \emph{ordered} species pair. This allows us to restrict, for example, H-C ordered pairs to a stricter cutoff than C-H ordered pairs. In such a case the C-centered ordered pairs can still be many-body with regard to H atoms out to the larger cutoff, but the H-C pairs are restricted to a smaller set, reducing computational cost. On the SPICE data set, we found the use of such reduced per-ordered-pair cutoffs for hydrogen to impose a negligible accuracy cost of less than 2 meV/\AA\ in validation force RMSE. Using these selected per-species-pair cutoffs (see section \ref{sec:trainingdetails}) reduces the number of ordered atom pairs by approximately $3\times$ in a liquid water system compared to using the maximum per-species-pair cutoff across all species. Because Allegro is linear-scaling in the total number of neighbor pairs \cite{musaelian2023learning} this correspondingly reduces the computational cost of the model.
\subsection{Optimized LAMMPS-Kokkos implementation}
LAMMPS is among the most popular molecular dynamics codes, thanks to its ease of use and performance \cite{LAMMPS}. In particular, LAMMPS has a state-of-the-art spatial decomposition algorithm for MD simulations on distributed memory hardware, and its scalability is well-proven on the world's largest supercomputers. We have chosen to implement an Allegro plugin to LAMMPS by carefully implementing the Allegro model to be compatible with C++ export through PyTorch's TorchScript JIT compiler, allowing us to call the TorchScript-compiled model from our plugin through the \texttt{libtorch} C++ API. Since Allegro is a strictly local model, it fits perfectly into the spatial decomposition concept of LAMMPS, thus LAMMPS will handle all inter-rank communication. Initially, the Allegro-LAMMPS interface let LAMMPS compute neighbor lists etc.\ on the CPU, and then copied the relevant data to the GPU before calling the PyTorch model. This was based on the assumption that Allegro would be far too slow, even on the GPU, for any CPU work or CPU-GPU memory transfers to appreciably affect the overall performance. With the optimizations of Allegro described in this paper, that assumption was no longer valid. LAMMPS has the ability to use the Kokkos performance portability library to accelerate its core functionality with GPUs. In the newest version of the Allegro-LAMMPS interface, it directly uses the positions and neighbor lists already on the GPU and preprocesses them for the Allegro model without ever copying any data to the CPU\@. This vastly accelerates any work that would previously have been done on the CPU \emph{and} eliminates any data transfer between the CPU and the GPU\@ (except when writing output to disk). Finally, it allows LAMMPS to employ CUDA-aware GPU-GPU MPI operations when available. In our initial benchmarks, we discovered that the performance fluctuated for a significant amount of time at the beginning of each simulation. Using NVIDIA Nsight Systems we determined the cause to be large deallocations and allocations of memory by the internal PyTorch memory handler whenever the shapes of the input tensors of the Allegro model changed. These shapes are determined by the number of atoms per GPU and their total number of neighbors, which fluctuate during an MD simulation. As a circumvention, we pad the Kokkos data structures from which the input tensors are created by 5\% whenever they need to be allocated. The extra memory is filled with edges between two ``fake'' atoms far apart. This largely eliminates internal PyTorch reallocations, as shown in figure~\ref{fig:padding}.

\section{How Performance Was Measured\label{sec:perf}}

\subsection{Scientific applications used to measure performance}\label{subsec:sciapp}

We measure MD performance on a varity of biomolecular systems. First, from the explicit all-atom water solvent \mbox{AMBER20} benchmark \cite{AMBER20Bench} we take the proteins Dihydrofolate Reductase (DHFR, 23k atoms) and human clotting factor IX (91k atoms), a cellulose sugar polymer (409k atoms), and the complete satellite tobacco mosaic virus (STMV, 1M atoms). At larger scale, we measure performance on a complete all-atom HIV capsid solvated in and containing water from \cite{yu2022strain}, which is a 44M atom assembly of protein subunits. MD simulations are performed with an Allegro model trained on approximately 1 million DFT reference calculations, using the recent SPICE data set \cite{eastman2023spice}, which aims to train MLIPs for biomolecular systems. The SPICE data set uses the $\omega$B97M-D3(BJ) functional, which is among the most accurate hybrid functionals for main-group chemistry benchmarks, and the def2-TZVPPD basis set, which was chosen to maximize basis set accuracy within computational budget constraints (for additional details, see \cite{eastman2023spice}). MLIPs come with an inherent trade-off between accuracy and efficiency even within a model class, where larger models are often more accurate, but simultaneously more computationally expensive. The Allegro model used here is a comparatively large and powerful model with 7.85 million weights (see section \ref{sec:trainingdetails}). Faster, albeit less accurate, models could certainly be trained. On a 55,353 frame hold-out test set of the SPICE data, we obtain a mean absolute error in the force components of 25.7 meV/\AA\ and an RMSE of 48.1 meV/\AA, demonstrating the high accuracy of the Allegro potential. We stress, however, that the goal of the present work is not to build the most powerful general-purpose potential for biomolecular simulations, but rather to demonstrate that Allegro provides a framework for large-scale biomolecular simulation due to its ability to retain the high accuracy of equivariant deep learning while being able to scale to large and long simulations. We expect that larger and more diverse quantum training data generated with higher levels of accuracy will continue to improve Allegro potentials. We further note that due to the strict locality, explicit long-range electrostatic interactions are straightforward to add to the Allegro potential, if they are required, following for example \cite{ko2021fourth}.

\subsection{Systems and Environment for Measurement of Performance}
All LAMMPS performance benchmarks were performed on the Perlmutter machine at NERSC\@. Perlmutter consists of 1536 nodes on the regular queue, each equipped with 4 NVIDIA A100, each with 40 GB of memory, and an AMD EPYC 7763 (Milan) CPU\@. The performance was measured using LAMMPS's profiling output, measuring timesteps per second and nanoseconds per day. All experiments were run with the Allegro code available at \url{https://github.com/mir-group/allegro} under git commit \texttt{cc76ba89a489e77329e69a2c5381fdf49bfc1060} and version 0.3.0. In addition, we used the NequIP code available at \url{https://github.com/mir-group/nequip} with version 0.6.0, git commit \texttt{671c369feba4e0140e50b9e34730c62e4023958a}, as well as \texttt{e3nn} with version 0.5.1 \cite{geiger2022e3nn}, PyTorch with version 1.11.0, and Python with version 3.9.16. The LAMMPS experiments were run with the LAMMPS code available at \url{https://github.com/lammps/lammps} under git commit \texttt{fce1f8e0af22106abece97c8099815e97c8980c6} with the \texttt{pair\_allegro} code available at \url{https://github.com/mir-group/pair_allegro},  git commit \texttt{e917edb940e1bfc429a2f3c18e1e1e243432118f}. The compilers and libraries used on Perlmutter were CUDA 11.7, cuDNN 8.7.0, GCC 11.2.0 and Cray MPICH 8.1.25. Due to issues with CUDA-aware MPI with the Cray MPICH library, LAMMPS was told to \emph{not} use CUDA-aware MPI through the \texttt{gpu/aware off} flag of the KOKKOS package, which may hurt scalability for the largest numbers of nodes. All jobs were launched with 4 MPI tasks per node (one per GPU) and \texttt{sbatch --gpu-bind=none}, with LAMMPS being responsible for assigning the GPUs.
\subsection{Measurement metrics}
The speed of the molecular dynamics simulations was measured in timesteps per second, which can be converted to nanoseconds per day of simulation time given a choice of timestep. In the beginning of a simulation, the PyTorch JIT compiler is still active while the system is also reaching an equilibrium number of neighbors per atom and atoms per GPU, thus the performance will be somewhat unstable. We thus first ran a fixed number of timesteps, typically 200, before measuring the performance. The peak performance was then measured during short, subsequent simulations.

\subsection{Training details}\label{sec:trainingdetails}

All training was performed using a single A100 GPU. We train in units of eV and Angstrom and use the FP64, FP32, TF32 configuration described above. We trained on 996,352 structures and use 55,353 as a validation set, as well as a remaining 55,353 as a separate test set. Before this split, we filter out all structures in SPICE v1.1.3 that contain any force component with an absolute value larger than 0.25 Ha/Bohr, which results in a total of 1,107,058 structures. Atom types in the model correspond one-to-one with chemical species. The Allegro models used two layers of 64 tensor features with a $\ell_{max}=2$, full $O(3)$ symmetry. The data set was re-shuffled after each epoch. We used a two-body latent MLP and later latent MLP with hidden dimensions [128, 256, 512, 1024] and [1024, 1024, 1024] respectively, both with SiLU nonlinearities. The embedding MLP was a linear projection. For the final edge energy MLP, we used a single hidden layer of dimension 128 and no nonlinearity. All four MLPs were initialized according to a uniform distribution of unit variance. Models were trained with a default radial cutoff of 4.0~\AA, which we found sufficient to give strong performance. We make use of the per-pair cutoffs outlined above, which were chosen based on radial distribution functions of the HIV capsid starting structure: H-H: 3.0~\AA, H-C: 1.25~\AA, H-O: 1.25~\AA, O-H: 3.0~\AA, where all others use 4.0~\AA. We stress that pair indices in Allegro are \emph{ordered}, and therefore an H-C cutoff of 1.25~\AA\ does not imply the same of C-H, which in our model uses the full 4.0~\AA\ cutoff. The interatomic distances were embedded in a trainable per-ordered-species-pair radial basis of 8 Bessel functions and a polynomial cutoff envelope function as specified in~\cite{musaelian2023learning}. The potential was trained using a force-only MSE loss function. The potential was trained with the Adam optimizer in PyTorch using default settings \cite{paszke2019pytorch}. Models were trained with a batch size of 16 and a learning rate of 0.001. For the checkpoint used for the test set evaluations, we reduced the learning rate after 119 epochs to 0.0005 and then train for an additional 23 epochs. The production MD was performed with an earlier checkpoint, with only minor differences in accuracy. We also used an exponential moving average on the network weights with a decay weight of 0.99, used for evaluation on the validation set and for the final model. We stopped training after approximately 7 days. We note that strong results can be obtained with less training time. We normalize the force targets by the maximum absolute force component computed over the training set. We also add a repulsive Ziegler-Biersack-Littmark (ZBL) term to the potential as a means to improve the stability of the potential.

\section{Performance Results}\label{sec:performanceresults}

\subsection{Stability and single-node performance}
Before running large-scale benchmarks of biomolecular systems, the  stability of the Allegro model must be verified. To this end, we performed long simulations of the solvated DHFR and factor IX proteins and measured as a function of time the root mean squared distance (RMSD) of the backbone atoms of the protein with respect to the initial structure. The results are shown in figure~\ref{fig:stability}, with the RMSD of both proteins being stable for the more than 3 ns of simulation conducted.

We then validate our mixed precision approach: table~\ref{tab:mixed-precision} shows that the use of mixed precision has no effect on the accuracy of the model, that our limited use of \texttt{float64} has no impact on speed, and that the careful use of \texttt{TensorFloat32} more than doubles the model speed, confirming our choice of a F64,F32,TF32 architecture. In particular, we note that the use of the tensor cores available on the NVIDIA A100 GPUs improved the performance by a factor of 2.7. Without \texttt{TensorFloat32}, the tensor cores, and thus a significant fraction of the GPU capacity, would be unused.

Finally, we verify the effect of padding the input tensors to the PyTorch model in figure~\ref{fig:padding}. The 5\% overallocation of memory clearly helps PyTorch's internal memory handler avoid unnecessary allocations and ensures smooth, stable performance. This both improves performance for shorter simulations and vastly simplifies the benchmarking task. The performance was measured on 1 node with 4 GPUs, simulating 100k atoms of water starting from an equilibrated structure.

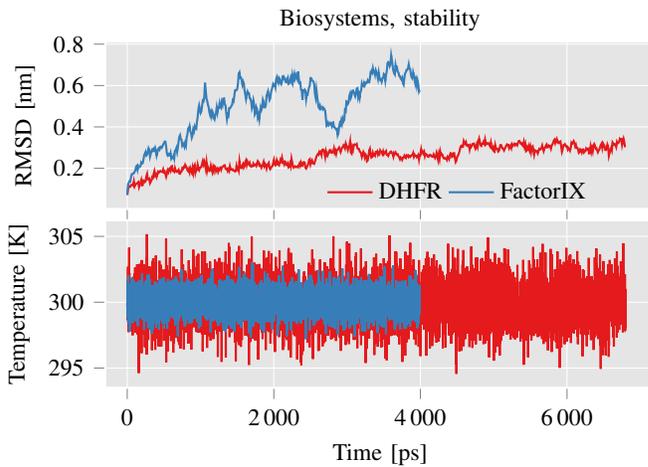
\begin{figure}[htbp]
        %\tikzexternalenable
        \begin{tikzpicture}[>=latex]
            \begin{groupplot}[
                group style={
                    group size=1 by 2,
                    xticklabels at=edge bottom,
                    xlabels at=edge bottom,
                    vertical sep=0.15cm
                },
                /tikz/thick,
                width=3.5in,
                height=1.5in,
                xmin=-300,
                xmax=7200, 
                xlabel={Time [ps]},
            ]
                \nextgroupplot[
                        ylabel={RMSD [nm]},
                        title={Biosystems, stability},
                        legend columns=2,
                        legend style={at={(0.9,0.1)},anchor=east},
                        title style={yshift=-0.2cm},
                    ]
                    \addplot+[] table[x index=1, y index=2] {data/JAC-rmsd.dat};
                    \addlegendentry{DHFR};
                    \addplot+[] table[x index=1, y index=2] {data/FactorIX-rmsd.dat};
                    \addlegendentry{FactorIX};
                \nextgroupplot[
                        ylabel={Temperature [K]},
                    ]
                    \addplot+[] table[x index=0, y index=1] {data/JAC-temp.dat};
                    \addplot+[] table[x index=0, y index=1] {data/FactorIX-temp.dat};
            \end{groupplot}
        \end{tikzpicture}
        %\tikzexternaldisable
        \caption{Top: RMSD of protein backbone atoms with regard to the initial structure. Bottom: stable temperature around thermostat setting of 300K.}\label{fig:stability}
\end{figure}

\begin{table*}[htb]
    \centering
    \caption{Test set RMSE comparison in [meV/\AA] of various mixed precision schemes. Models were trained from scratch on the water dataset from \cite{zhang2018deep}. Speed is measured in MD started from $3072 \text{ atoms } =16\times192$ atom of liquid water test frame running on one 80GB A100.}
    \label{tab:mixed-precision}
    \begin{tabular}{|r|rrrrr|}
        \hline
        \multicolumn{1}{|c|}{Precision} & \multirow{2}{*}{F32,F32,TF32} & \multirow{2}{*}{F32,F32,F32} & \multirow{2}{*}{\textbf{F64,F32,TF32}} & \multirow{2}{*}{F64,F32,F32} & \multirow{2}{*}{F64,F64,F64} \\
        Final, Weights, Compute & & & & & \\
        \hline
        Liquid Water & 29.0 & 28.8 & 29.1 & 28.6 & 28.7  \\
        Ice Ih (b)   & 30.5 & 30.3 & 30.7 & 30.1  & 30.5 \\
        Ice Ih (c)   & 20.9 & 20.7 & 21.0 &  20.7& 20.8 \\
        Ice Ih (d)   & 17.9 & 17.7 & 18.0 &  17.7& 17.7 \\
        \hline
        speed vs.~F64,F32,TF32 & $0.98\times$ & $0.37\times$  & $1\times$ & $0.37\times$ & $0.26\times$ \\
        \hline
    \end{tabular} 
\end{table*}

\begin{figure}[htbp]
        %\tikzexternalenable
        \begin{tikzpicture}[>=latex]
            \begin{axis}[
                    thick,
                    width=3.0in,
                    height=2.0in,
                    ylabel near ticks,
                    xlabel={Step},
                    ylabel={Speed [timesteps/s]},
                    legend style={at={(0.92,0.2)},anchor=east},
                    title={Effect of padding},
                    title style={yshift=-0.75cm},
                    ymax=5.3,
                ]
                \addplot+[] table[x index=0, y index=7] {data/nopadperf.dat};
                \addlegendentry{Without padding};
                \addplot+[] table[x index=0, y index=7] {data/padperf.dat};
                \addlegendentry{With padding};
            \end{axis}
        \end{tikzpicture}
        %\tikzexternaldisable
        \caption{The effect of padding all input tensor to the PyTorch model, which avoids unnecessary internal reallocations by the PyTorch memory handler. This stabilizes the performance much faster.}\label{fig:padding}
\end{figure}
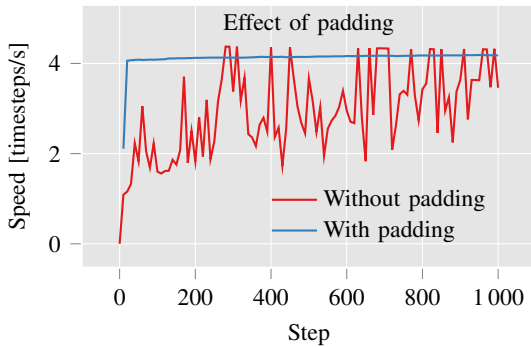

\subsection{Scalability\label{subsec:scalability}}
We measured the scalability and performance on a variety of systems and sizes relevant for biological simulations. First and foremost, we simulate the systems described in section~\ref{subsec:sciapp}, ranging in size from 20k to 44M atoms. For each system, we examined strong scaling from the minimum number of nodes required up to 1280 nodes (5120 GPUs). Similarly, we performed benchmarks with the same potential on water systems, which are both highly relevant for biological applications and better suited to ``custom'' system sizes for weak scaling experiments. We performed both strong scaling tests ranging from 100k to 100M atoms and weak scaling experiments ranging from 25k to 100k atoms \emph{per node} for the water system, replicated isotropically from a 192-atom unit cell.

The strong scaling results are shown in figure~\ref{fig:strongscaling}, with the number of nodes increased for each system until the performance saturated. Allegro achieved performance in excess of 100 timesteps/s for all systems up to 1M atoms, both water and biological. This performance was consistently achieved when the number of atoms per GPU dropped below 500, which limits the saturation of the GPU. When using fewer GPUs, each GPU was fully saturated and the scaling was near-linear. This demonstrates the efficiency of Allegro in taking advantage of the massive parallelism offered by GPU resources. With deep equivariant models requiring more work per atom than classical force fields, the model evaluation on the GPU is by far the dominant bottleneck, and its efficiency is paramount to the overall speed of the simulation. The dense linear algebra operations comprising the tensor product and scalar track MLPs for every ordered neighbor pair in an Allegro model expose many levels of parallelism and ensure high efficiency even when running on leadership computing facilities with few atoms per GPU. In contrast, classical force fields commonly only parallelize over atoms and thus require hundreds of thousands of atoms per GPU to reach peak performance and maintain linear scaling. For the biggest systems, the peak performance was 36.3 and 4.32 timesteps/s for 10M and 100M atoms of water, respectively, and 106, 23.0 and 8.73 timesteps/s for the STMV, 10\(\times\)STMV and atom capsid with 1M, 10M and 44M atoms, respectively. The STMV and 10\(\times\)STMV numbers can be directly compared with the performance numbers of 268 and 24 timesteps/s with classical force fields in the widely used Desmond MD code~\cite{desmond2021}. It should be noted that Desmond only supports single-GPU execution, but this demonstrates how the scalability of Allegro enables it to reach practical performance comparable to that of current, much less accurate methods through the use of current supercomputing hardware. Finally, we can favorably compare the speeds we achieve simulating the 44M atom HIV capsid---3.9-8.7 timesteps/s on 512-1280 nodes---to a previous effort to simulate a 62M atom HIV capsid at quantum accuracy in~\cite{schade2022towards}, which achieved 0.0005 timesteps/s on 384 nodes.

Figure~\ref{fig:weakscaling} shows the weak scaling results for water, where the number of atoms per node was kept approximately constant while the number of nodes increased. We scaled the water MD simulations from 1 to 1280 nodes, with 25k, 50k, 75k, and 100k atoms per node. Excellent weak scaling in excess of 70\% is achieved, in particular for the larger system sizes. For the smaller system sizes, the performance degrades for the largest sizes due to the communication overhead. With 100k atoms per node, we encountered MPI errors in the jobs ran on 128 and 256 nodes, and these data points were omitted.

\begin{figure}[htbp]
        %\tikzexternalenable
        \begin{tikzpicture}[>=latex]
            \begin{groupplot}[
                /tikz/thick,
                group style={
                    group size=1 by 2,
                    vertical sep=0.3cm,
                    xlabels at=edge bottom,
                    ylabels at=edge left,
                    xticklabels at=edge bottom,
                    horizontal sep=3.0cm,
                },
                xlabel={\# nodes},
                width=2.7in,
                height=2.0in,
                xmode=log,
                ymode=log,
                xmin=0.8,
                xmax=1500,
                ymax=300,
                ylabel near ticks,
                ylabel={Speed [timesteps/s]},
                legend style={at={(1.0,0.5)},anchor=west},
                log basis x=2,
                log x ticks with fixed point,
                title style={yshift=-0.75cm},
            ]
            \nextgroupplot[
                title={Biosystems},
            ]
                \addplot+[mark=*] table[y index=3] {data/dhfr.dat};
                \addlegendentry{DHFR};
                \addplot+[mark=triangle*] table[y index=3] {data/factorix.dat};
                \addlegendentry{FactorIX};
                \addplot+[mark=diamond*] table[y index=3] {data/cellulose.dat};
                \addlegendentry{Cellulose};
                \addplot+[mark=square*] table[y index=3] {data/stmv.dat};
                \addlegendentry{STMV};
                \addplot+[mark=square*, mark options={rotate=45}] table[y index=3] {data/bigstmv.dat};
                \addlegendentry{10\(\times\)STMV};
                
                \pgfplotsset{cycle list shift=6}
                \addplot+[mark=pentagon*] table[y index=3] {data/capsid.dat};
                \addlegendentry{Capsid};
            \nextgroupplot[
                    title={Water},
                ]
                \addlegendimage{empty legend};
                \addlegendentry{\hspace{-0.65cm}\# atoms};

                \addplot+[mark=*] table[y index=3] {data/waterstrong1e5.dat};
                \addlegendentry{$10^5$};
                \addplot+[mark=square*] table[y index=3] {data/waterstrong1e6.dat};
                \addlegendentry{$10^6$};
                \addplot+[mark=triangle*] table[y index=3] {data/waterstrong1e7.dat};
                \addlegendentry{$10^7$};
                \addplot+[mark=diamond*] table[y index=3] {data/waterstrong1e8.dat};
                \addlegendentry{$10^8$};
            \end{groupplot}
        \end{tikzpicture}
        %\tikzexternaldisable
        \caption{Strong scaling performance for biomolecular systems and water from 1 to 1280 nodes. DHFR, factor IX, cellulose, STMV, 10\(\times\)STMV, and capsid contain 24k, 91k, 409k, 1M, 10M, and 44M atoms, respectively. We achieve near-linear scaling until the performance reaches 100 timesteps/s, when the GPU saturation decreases and the communication overhead becomes noticeable.}\label{fig:strongscaling}
\end{figure}
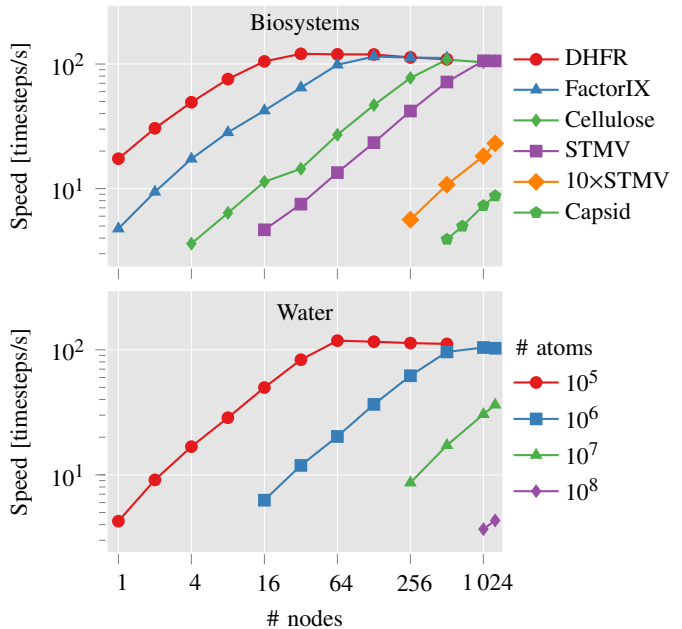
\begin{figure}[htbp]
        %\tikzexternalenable
        \begin{tikzpicture}[>=latex]
            \begin{axis}[
                    thick,
                    width=3.0in,
                    height=2.5in,
                    xmode=log,
                    xmin=0.8,
                    xmax=1500,
                    ylabel near ticks,
                    xlabel={\# nodes},
                    ylabel={Speed [timesteps/s]},
                    legend style={at={(0.92,0.5)},anchor=east},
                    legend columns=2,
                    log basis x=2,
                    log x ticks with fixed point,
                    title={Water, weak scaling},
                    title style={yshift=-0.75cm},
                ]
                \addlegendimage{empty legend};
                \addlegendentry{\hspace{-0.65cm}\# atoms/node};
                \addlegendimage{empty legend};
                \addlegendentry{};
                \addplot+[mark=*] table[y index=3] {data/waterweak25k.dat};
                \addlegendentry{25k};
                \addplot+[mark=square*] table[y index=3] {data/waterweak50k.dat};
                \addlegendentry{50k};
                \addplot+[mark=triangle*] table[y index=3] {data/waterweak75k.dat};
                \addlegendentry{75k};
                \addplot+[mark=diamond*] table[y index=3] {data/waterweak100k.dat};
                \addlegendentry{100k};
            \end{axis}
        \end{tikzpicture}
        %\tikzexternaldisable
        \caption{Weak scaling of water from 1 to 1280 nodes, with different system sizes per node. With only 25k atoms per node, the performance eventually drops with communication becoming an overhead, while the larger sizes show excellent scaling. For the largest size, two of the calculations failed with MPI errors and have been omitted.}\label{fig:weakscaling}
\end{figure}
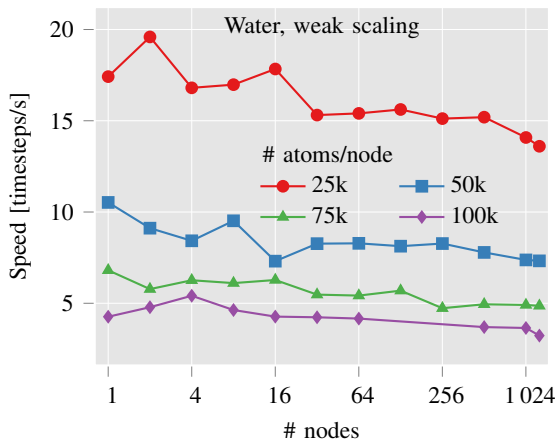

\section{Implications} 
Understanding the time evolution of complex systems containing tens and hundreds of millions of atoms is important for a wide range of heterogeneous and disordered materials systems and is particularly necessary in the field of biology, where even the simplest lifeforms and their building blocks possess these size magnitudes. At the same time, very high accuracy is required for faithful simulations of kinetics of chemical and biological processes. Equivariant models are now widely accepted to be unique in their leading ability to accurately capture quantum many-body interactions, surpassing all earlier empirical and machine learning potentials. This work effectively connects the highest accuracy currently achievable for interatomic interaction models with the extreme scalability afforded by leadership GPU computing. As such, it establishes the new state of the art for molecular dynamics and opens doors to simulating previously inaccessible systems. 

For the first time, we achieve large-scale molecular dynamics simulations of complex biological systems, entirely with machine learning interatomic potentials at quantum accuracy, which represents a new stage in biomolecular simulations. The specific biological systems were chosen to demonstrate the achievable accuracy at scale and complexity. However, the Allegro architecture can be used to simulate dynamics of any atomistic structure, e.g. polycrystalline and multi-phase composites, diffusion in glasses, polymerization and catalytic reactions, etc. The wide impact of the approach is also evident already by the rapid pace of adoption of equivariant interatomic potentials models in the research community. 

The combined use of the PyTorch and Kokkos performance portable libraries allows deployment of our state-of-the-art equivariant model architecture on a wide range of hardware computing architectures, including CPUs, and NVIDIA, AMD and Intel GPUs which are powering leadership-class resources. In the near future it will be possible to deploy Allegro on newer computing resources at even larger scale than what was shown in this work. By implementing our models in LAMMPS, we unlock access to its wide ecosystem of MD and enhanced-sampling simulation methods and analysis tools. These integrations, and the Allegro framework needed to train models to take advantage of them, are all open-source and designed for the use of the community.

Our work demonstrates the advantages of high-capacity equivariant Allegro models in accurately learning forces across the entire SPICE dataset of over 1 million structures of drug-like molecules and peptides. This data scale implies the promise of learning the entire sets of inorganic materials and organic molecules far more accurately than previously attempted, which would open the prospects of fast exascale simulations of unprecedentedly wide ranges of materials systems.
Recently we demonstrated that it is possible to efficiently quantify uncertainty of deep equivariant model predictions of forces and energies \cite{zhu2022fast} and use it to perform active learning for automatic construction of training sets. Natural adaptation of Gaussian mixture models in Allegro will open the possibility of large-scale uncertainty-aware simulations using a single model, as opposed to ensembles.

Finally, the major implication of the demonstrated accuracy of equivariant models is the urgent need to improve the accuracy and efficiency of the quantum electron structure calculations that are used as reference to train MLIPs, as this now becomes the major accuracy bottleneck in computational chemistry, biology and materials science.

\section{Acknowledgement}

We thank Peter Eastman, Stan Moore, Rahulkumar Gayatri, Kyle Bystrom, Blake Duschatko, David Clark, Lixin Sun, Mordechai Kornbluth, and Cameron Owen for useful discussions. We thank Gregory Voth for sharing structure files for the HIV capsid. 

B.K. acknowledges support from Bosch Research. S.B. was supported by DOE Office of Basic Energy Sciences Award No. DE-SC0022199 and Department of Navy award N00014-20-1-2418 issued by the Office of Naval Research. A.J. was supported by the NSF through the Harvard University Materials Research Science and Engineering Center Grant No. DMR-2011754. A.M. is supported by DOE, Scientific Computing Research, Computational Science Graduate Fellowship under Award Number DE-SC0021110. Computing resources were provided by NERSC on Perlmutter through the ERCAP0024206 allocation and by the Harvard University FAS Division of Science Research Computing Group.
\def\url#1{}
\def\doi#1{}

\bibliographystyle{IEEEtran}
\bibliography{main}

\end{document}